\documentclass{aa}  
\usepackage{graphicx}
\usepackage{txfonts}
\begin{document}
   \title{White Dwarf cooling Sequences, II: luminosity functions}


   \author{P. G. Prada Moroni
          \inst{1}\fnmsep\inst{2}
          \and
           O. Straniero\inst{3}
          }

   \offprints{P. G. Prada Moroni}

   \institute{Dipartimento di Fisica ``E. Fermi'', University of Pisa,
              largo B. Pontecorvo 3, 56127 Pisa (Italy)\\
         \and
              INFN,  largo B. Pontecorvo 3, 56127 Pisa (Italy)\\
              \email{prada@df.unipi.it}
         \and
              INAF - Osservatorio Astronomico di Collurania, via Maggini, 64100 Teramo\\
              \email{straniero@te.astro.it}
             }

   \date{Received September 15, 1996; accepted March 16, 1997}

 
  \abstract
  {Given the
   importance of white dwarfs (WDs) in many fields of modern astrophysics, the precise knowledge of 
   the actual degree of accuracy  of the associated theoretical predictions is a primary
   task. In the first paper of a series dedicated to the modeling of
    WD structure and evolution we discussed
   the limits of the available theoretical studies of cooling sequences.}
   {In the present work we extend this analysis to isochrones and 
   luminosity functions of WDs belonging to old stellar systems, like globular or old disk 
   clusters. The discussion is focused on the most common DA,
   those with a CO core and an H-rich envelope.}
   {We  discuss, in particular, the variation of the age derived from the observed WD sequence
   caused by different assumptions about the conductive opacity 
   as well as that induced by changing the carbon abundance in the core.} 
   {The former causes a global uncertainty of the order of 10\% and the latter 
   of about 5\%. We discuss different choices of the initial-to-final mass relation, 
   which induces an uncertainty of 8\% on the GC age estimate.}  
   {}


   \maketitle
%

\section{Introduction}
   \begin{figure*}[ht]
   \centering
   \includegraphics[width=\columnwidth]{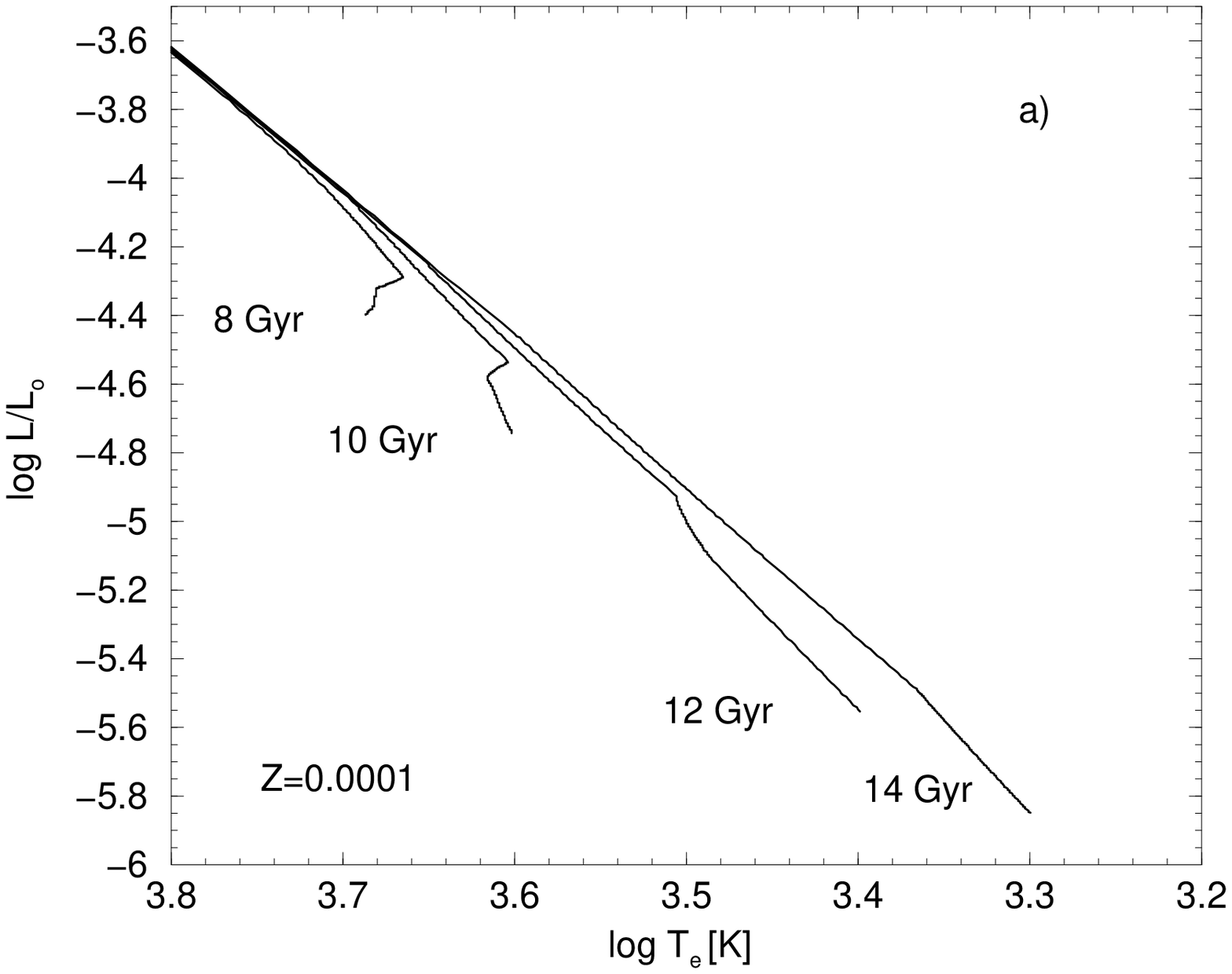}
   \includegraphics[width=\columnwidth]{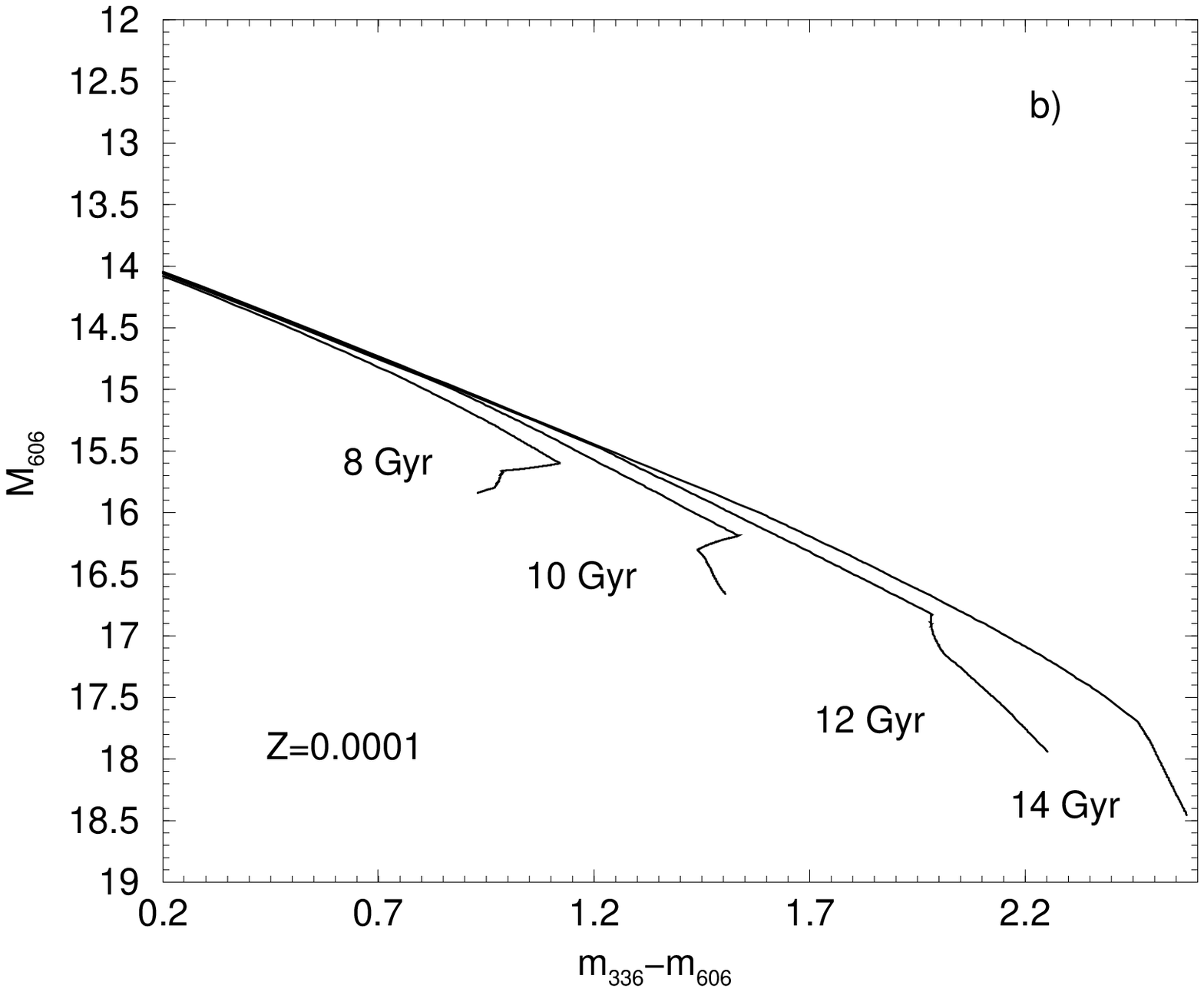}
   \includegraphics[width=\columnwidth]{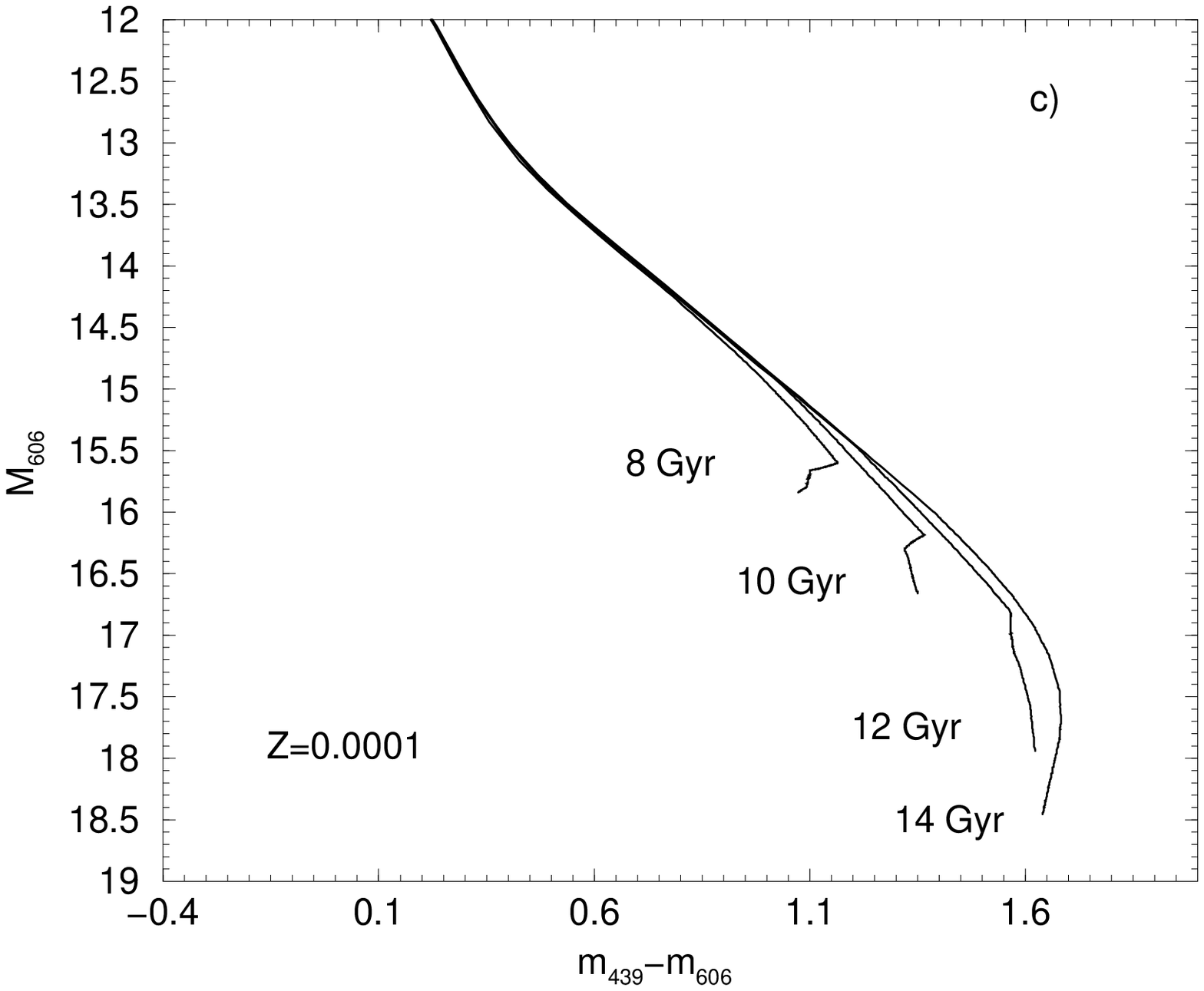}
   \includegraphics[width=\columnwidth]{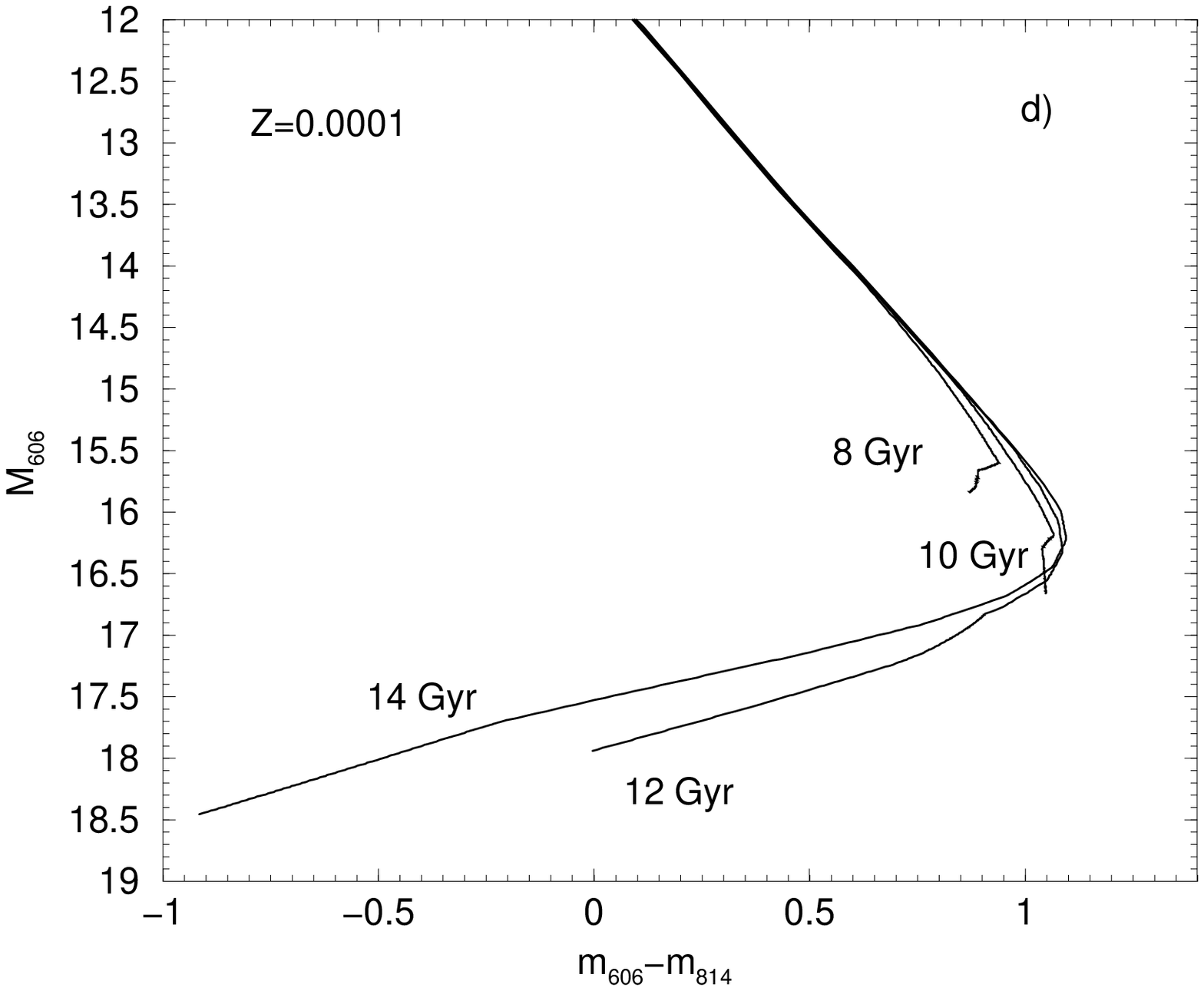}
   \caption{WD isochrones for ages in the interval 8-14 Gyr in the
   theoretical plane, i.e. $log L/L_{\sun}$ vs. $log T_e$ (panel a); V(606)
   vs. U(336)-V(606) (panel b); V(606) vs. B(439)-V(606) (panel c); V(606)
   vs. V(606)-I(814) (panel d). We adopted the HST/ACS transmission curves 
   $F336W$ (broad $U$), $F439W$ (broad $B$), $F606W$ (broad $V$) and
   $F814W$ (broad $I$).}
    \label{isocrone}
    \end{figure*}
  In the last decade, impressive improvement in observational techniques
  made available a significative sample of cold and faint
  white dwarfs (WDs) both in the field (Bergeron, Leggett \& Ruiz \cite{bergeron01})
  and in stellar clusters (Paresce, De Marchi \& Romaniello \cite{paresce},
  Cool, Piotto \& King  \cite{cool}, Renzini et al. \cite{renzini}, 
  Richer et al. \cite{richer}, Von Hippel \& Gilmore \cite{vonhippel}, 
  Zoccali et al. \cite{zoccali}, Hansen et al. \cite{hansen02}, 
  De Marchi et al. \cite{demarchi}, Monelli et al. \cite{monelli}).
  The growing amount of data prompted a renewed interest in theoretical
  studies of WD evolution, particularly for the latest stages
  when the core crystallizes and the stars approaches
  the Debye regime (Benvenuto \& Althaus \cite{benvenuto}, Hansen \cite{hansen},
  Chabrier et al. \cite{chabrier}, Salaris et al. \cite{salaris},
  Prada Moroni \& Straniero \cite{prada}, hereafter Paper I).

  Concerning the core composition, the theory of stellar evolution predicts three kinds of WDs:
  helium, carbon/oxygen (C-O), oxygen/neon/magnesium. The most common are,
  by far, those with a C-O core, which represent the final stage of the evolution 
  of single (non-interacting)
  stars with mass lower than about 7 M$_{\odot}$ (M$_{up}$)
  and a lifetime shorter than the age of the Galaxy. 
  The envelope composition is also very important for the evaluation of the 
  cooling timescale.
  Mass loss, particularly during the AGB and post-AGB,
  may substantially erode the envelope. 
  Moreover, the H-rich envelope can be substantially reduced 
  in the born-again evolution scenario (Iben et al. \cite{iben83}, 
  Iben \& MacDonald \cite{iben95}, Herwig \cite{herwig05}).
  Nevertheless, the large majority of the
  observed WDs show an hydrogen-rich atmosphere
  (DA WDs), while only 20\% of them are hydrogen-deprived (DB WDs, 
   see e.g. Koester \& Chanmugam \cite{koester}).
   The present study deals with the most common WDs, namely the DA type having a C-O core.
  
  Here we are interested in the use of WDs as cosmic chronometers (Schmidt \cite{schmidt})
  and standard candles (Fusi Pecci \& Renzini \cite{fusipecci}).
  Due to the intrinsic faintness of WD sequences in stellar clusters, 
  only very recently have these tools become useful.
  Their application to the nearest globular clusters (GCs) 
  provides age and distance estimates that are independent of those 
  obtained by means of the
  classical methods (those based on the main sequence and horizontal branch stars),
   thus allowing us to further constrain the lower
  limit of the Universe's age (Hansen et al. \cite{hansen02}, 
  De Marchi et al. \cite{demarchi}).
  One should, however, check the reliability of the
  theoretical cooling time predictions before adopting WDs as cosmic clocks.
  Notwithstanding the significant advances in the computation of
  the WD structure and evolution, there is still a
  sizeable uncertainty in the prediction of the cooling times
  of the old WDs (Montgomery et al. \cite{montgomery}, Salaris et al.\cite{salaris},
  Fontaine et al. \cite{fontaine}, Paper I).
  At low luminosity ($logL/L_{\sun}\sim -5.5$)
  the differences among the most recent models available in the literature
  reach about 4 Gyr (see e.g. fig. 1 in Paper I).
  In Paper I we analyzed some of the main
  sources of uncertainty in the evolution of a 0.6 $M_{\sun}$  
  C-O WD. In particular,
  we discussed uncertainties due to both the WD progenitor history and the
  adopted input physics.

  The evolution of a WD can be roughly described as a cooling process
  where the luminosity is supplied by the heat content of the
  internal matter; thermonuclear energy generation is negligible,
  except during the very early phase. Thus, the evolution results
  from the balance between the thermal energy stored in the C-O ions constituting
  the core (about 98\% of the WD mass) and the energy transport through
  the He-rich mantel and the H-rich envelope, the most opaque region of the star.

  In Paper I, we showed that the factors that most severely limit the accuracy 
  of theoretical predictions are the assumed conductive opacity in the partially
  degenerate regime, a condition that is usually fulfilled in the outer layer of the 
  core and in the He-rich mantel, and the amount of carbon and oxygen in the core.
  At $logL/L_{\sun}\sim -5.5$, these two sources produce a 
  total uncertainty on the predicted cooling age  of about 27\%.
  However, the analysis reported in Paper I gives only a rough idea 
  of the uncertainty in dating GCs
  with WDs, while a more precise estimate of this uncertainty
  should be based on WD luminosity functions instead of tracks of given mass, 
  since this is the best WD chronometer.
  For this work we have calculated several grids of models of C-O WDs with a 
  hydrogen rich atmosphere, with mass in the range 0.5-0.9 $M_{\sun}$, under
  different assumptions of the conductive opacity and the C-O profile. 
  The related isochrones and luminosity functions   
  (LFs) are compared to reveal the main uncertainty factors.
\section{Dating WDs in old stellar clusters: the piling up in the luminosity function}
  For old stellar systems, such as GCs, the best chronometer relying 
  on WD evolution is their luminosity function
 instead of the related isochrones. 
  Figure \ref{isocrone} shows WD isochrones for ages in the interval 8-14 Gyr
  in the theoretical plane ($log L/L_{\sun}$ vs. $log T_e$, panel a) 
  and in three CM diagrams for different HST/ACS pass-bands, $F336W$ (broad $U$),
  $F439W$ (broad $B$), $F606W$ (broad $V$) and $F814W$ (broad $I$). 
    In the theoretical plane (panel a), there is a clear and unambiguous 
  age dependent feature: the blue-turn of the WD isochrone, which plays
  the same role of clock pointer as the turn-off point in the main sequence
  phase. The luminosity of the blue-turn depends only on the evolutionary 
  time scales, both of the WDs and of their progenitors. 
  On the other hand, on the observational CM diagram, the situation is less 
  clear, in fact the morphology of the old ($>$10 Gyr) WD isochrones 
  significantly depends on the adopted pass-bands, particularly the luminosity 
  of the blue-turn. In the $V$(606) vs. $V$(606)-$I$(814) diagram,
  commonly used in observations of cool WDs, the blue-turn is almost
  insensitive to the age for isochrones older than about 10 Gyr. 
  This behavior is a consequence of the strong depletion of the IR flux 
  in the emerging spectrum of cool ($T_e <$  5000 K) DA WDs due to the
  collision induced absorption (CIA) of H$_2$-H$_2$ 
  (see e.g. Bergeron, Saumon \& Wesemael \cite{BSW}).  
  Since the latter blue shift is   
   an atmospheric effect that reflects the departure of
  the emergent spectrum from the black body appearance, its extension 
  depends sensitively on the wavelength of the transmission curves of the 
  filter adopted in the observations. The blue shift will be much larger 
  in the $V$-$I$ color than in the  $B$-$V$, because the CIA is the main opacity 
  source in the IR region of the spectrum. Such an occurrence explains why 
  in the $V$ vs. $V$-$I$ plane the blue-hook luminosity of old WD
  isochrones is essentially insensitive to the age and the blue-tail of the 14 Gyr 
  WD isochrone is brighter than the 12 Gyr one.
  The degeneracy of the position of the blue-turn with age and the 
  necessity to accurately model the CIA opacity sources in the 
  WD model atmospheres severely reduce the effectiveness of the WD isochrones
  in dating old stellar systems such as galactic GCs.
   \begin{figure}
   \centering
   \includegraphics[width=\columnwidth]{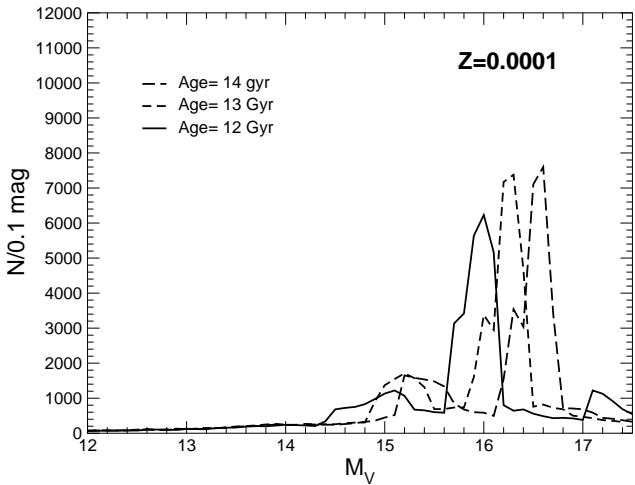}
   \caption{Theoretical LFs for 12 (solid line), 
    13 (dashed line) and 14 Gyr
   (long-dashed line).  Each magnitude bin is $\Delta$M$_V$=0.1 mag. The 
   total number of WDs is 10000.}
    \label{LF_age}
    \end{figure}
  A much firmer estimate relies on star counts, i.e. luminosity functions, 
  which directly reflect the evolutionary time scales.

  Figure \ref{LF_age} shows the WD 
  LFs for 12, 13, 14 Gyr.
  These and the following WD LFs have been computed with
  Monte Carlo simulations that distribute 50000 WDs along the related isochrones. 
  This figure clearly illustrates the potential of WD LFs in dating old stellar systems.
  Note that the peak of the luminosity 
  function shifts by about 0.3 mag per Gyr in the $V$ band, while the 
  main sequence turn-off luminosity, the clock classically
  used to date stellar clusters, shifts only by about 
  0.1 mag. This implies that the WD age estimate is significantly 
  less affected by the uncertainty on the distance
  modulus, which represents the main source of error in dating GCs with the method
  based on the turn off luminosity.  
  Let us briefly recall the main WD evolutionary phases in order 
  to explain the principal features in the WD isochrones and 
  LFs. 
  The evolution of WDs is initially characterized by a 
  rapid drop of the luminosity and of the central temperature. The main 
  energy loss characterizing this initial phase is the emission of neutrinos. 
  The corresponding zone of the isochrone closely follows the track 
  of a WD of fixed mass. 
  Then, as a WD cools, the Coulomb interaction between ions 
  becomes progressively greater and leads to the crystallization
  of carbon and eventually oxygen (Abrikosov \cite{abrikosov};
  Kirzhnits \cite{kirzhnits}; Salpeter \cite{salpeter}). 
  Due to the latent heat released by the liquid-solid transition, 
  the cooling rate slows temporarily (D'Antona \& Mazzitelli
  \cite{dantona}, Fontaine et al. \cite{fontaine}).   
  At the same time, the external convective zone penetrates 
  the region where thermal conduction by degenerate electrons 
  is very efficient. Such an occurrence (known as {\it{convective coupling}}),
  initially produces a further slowing down of the cooling timescale  
  (see e.g. Chabrier et al. \cite{chabrier}, Fontaine et al. \cite{fontaine}).
  The steep increase in the WD luminosity function is the effect of 
  such a decrease in the cooling rate.   
  Finally, the WD enters 
  the Debye regime, where the heat capacity decreases as $T^3$, thus
  depleting the WD of its main energy reservoir. As a consequence its luminosity 
  quickly drops. 
  
  The morphology of the luminosity function reflects  
  the variations of the cooling timescale:
  the LF is generally sparse at the brightest magnitudes and presents a steep rise,   
  followed by a sharp cutoff, in the faintest part (see
  e.g. Fontaine et al. \cite{fontaine}).
   
  The best way to date stellar systems through WDs is the analysis 
  of their LF.
\section{The model}
   \begin{figure}
   \centering
   \includegraphics[width=\columnwidth]{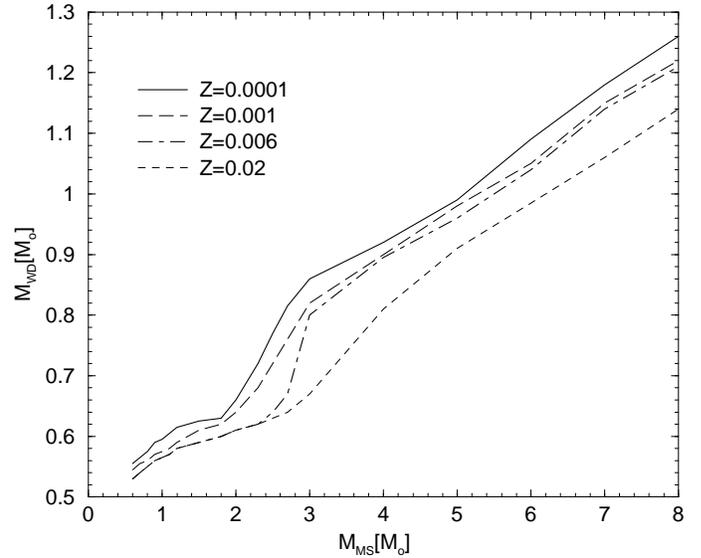}
   \caption{Initial-to-final mass relationship for the four labeled
 metallicities from our stellar models. See text for details.}
    \label{Mwd_Mms_Z}
    \end{figure}
  The present WD theoretical tracks have been computed
  with the FRANEC (Chieffi \& Straniero \cite{chieffi}), a full
  evolutionary code that adopts the Henyey method and that
  is able to model the whole evolution of stars from the hydrostatic pre-main
  sequence phase up to the WD stage. 
\subsection{Progenitors and starting model}
  To provide the starting model for the WD evolution, 
  we started from the ZAHB (Zero Age Horizontal Branch)
  model. The first model consists of a helium core (99\% of the total mass) 
  and a thin hydrogen rich envelope (Y=0.24 Z=10$^{-4}$). 
  Then, we followed the evolution at constant mass, through the core He-burning. 
  Due to the reduced envelope, at the central helium exhaustion the model rapidly moves 
  towards the blue region of the HR diagram. During this transition 
  the He- and the H-burning shells are eventually active. 
  As discussed in Paper I, the model settles on the cooling 
  sequence when the masses of  both the He-rich and the H-rich layers 
  are reduced to certain critical values and the thermonuclear burning dies down.  
  
  We are aware that in the real world a 
  DA C-O WD of a given mass can be formed through several different evolutionary paths, 
  depending on the mass loss rate and on the initial chemical
  composition. This was
  one of the main points discussed in Paper I, where we studied three extreme cases, 
  and we showed that the internal structure of the resulting WD 
  is poorly affected by the different evolutionary histories 
  (see e.g. Fig. 2 and 3 of Paper I). Moreover, the computation of the 
  evolution through the thermally pulsing asymptotic giant branch (TP-AGB), 
  although more realistic, is very time consuming. 
  Thus, the method we have adopted represents a useful way to provide realistic
  WD starting models consistent with the evolution of the progenitors. 
  In particular the chemical abundance profiles of carbon and oxygen in the core
  is largely independent of the evolutionary history. 
  Obviously, the computation through the TP-AGB evolution should be
  preferred if the initial, hot phase of WD evolution is analyzed. In 
  such an early phase of the cooling the outer layers still bear the 
  record of the complex chemical abundance profile left by TP-AGB evolution,
  which is subsequently modified by diffusion. 
  The metallicity of the progenitors determines the maximum value of the 
  envelope mass, but this quantity does not significantly affect the cooling timescale
  (see paper I). 
  The only important quantity
  is the relation between the progenitor mass and the WD mass (initial-to-final mass relation)
  which is needed to correctly evaluate the time spent before the cooling sequence and,
  in turn, correctly locate a WD with a certain mass  
  on the HR diagram (see 2.3).
  We stop the progenitor evolution when the newly formed WD reaches 
  log L/L$_{\odot} \sim $ 0. At this point, the helium and 
  hydrogen in the envelope are fully sorted, in order 
  to reproduce the observed chemical stratification  
  of the atmospheres of DA WDs. Several studies have indeed shown 
  that in such a compact stars, where the surface gravity $g$ is of the order of 
  $\sim 10^{8}-10^{9}$ cm sec$^{-2}$, the gravitational settling is efficient 
  enough to account for the monoelemental feature of 
  WD spectra (Fontaine \& Michaud \cite{fontaine79}, Muchmore \cite{muchmore}, 
  Paquette et al. \cite{paquette}). 
  The model obtained with this method is our starting point for the computation of the
  cooling sequence. 
\subsection{Cooling tracks}
  Let us recall the major features of the present models;
  for a detailed description see Paper I.
  We put considerable effort into updating the main input physics, namely the
  equation of state (EOS) for a high-density plasma, the radiative and conductive 
  opacity and the model atmospheres.   
  We updated and extended the EOS for a fully ionized plasma originally 
  described by Straniero (\cite{straniero88}). In particular, we have 
  adopted the most recent results for the electrostatic corrections both 
  in the liquid and in the solid phase (see Paper I for references).
  Concerning the Rosseland radiative opacity, we adopted the OPAL 
  (Iglesias \& Rogers \cite{iglesias}) for high temperature (log T[K] $>$ 4.0) 
  and the Alexander \& Ferguson (\cite{alexander}) molecular opacities for
  the low temperatures (log T [K] $<$ 4.0). 
  
  As we have shown in Paper I, 
  for the cooling evolution of old WDs, the treatment of electron conduction
  in the partially degenerate regimes typical of the outer layers is complex. 
   We have computed sets of theoretical cooling sequences and 
  the related isochrones adopting the conductive opacity by 
  Hubbard \& Lampe (\cite{hubbard}, hereafter HL69), by Itoh and coworkers 
  (Itoh et al. \cite{itoh83}, Mitake et al. \cite{mitake} and 
  Itoh et al. \cite{itoh93}, hereafter I93), Potekhin (\cite{potekhin},
  hereafter P99, see also Potekhin et al. \cite{potekhin2}), and
  combinations. The description of the different assumptions and 
  the related effects on the cooling sequences of WDs is given in section 5.

  The computations of reliable cooling models rely on the choice of
  appropriate boundary conditions, namely the pressure at the base 
  of the photosphere. As shown by Fontaine et al. (\cite{fontaine1})
   and then confirmed by successive studies
  (see e.g. Tassoul, Fontaine \& Winget \cite{tassoul}, Hansen \cite{hansen})
  it is important to use a detailed model atmosphere  
  to fix the surface boundary conditions.
  It is well known that one of the most important events in the
  evolution of old WDs, the coupling between the outer convective envelope
  and the isothermal degenerate core (see e.g. Chabrier et al. \cite{chabrier}, 
  Fontaine et al. \cite{fontaine}) is quite sensitive to the details of
  atmosphere stratification, in particular a gray approximation overestimate
  the inward penetration of the convection. The quoted coupling initially produces
  a sizeable slow down in the cooling evolution followed by a rapid drop in the
  luminosity. Since the occurrence of the convective coupling, the core,
  which is essentially the energy reservoir of the star, and the atmosphere, where
  the energy is radiated in space, are directly coupled.
  The pressure at the outer boundary, namely the layer where the optical
  depth $\tau=$ 50, has
  been obtained by interpolating on a grid of model atmospheres (Bergeron \cite{bergeron})
  which cover a range of surface gravities between 10$^{7.5}$-10$^{9}$
  cm/s$^2$ and effective temperatures between 1500-100000 K.  
  A detailed
  treatment of the atmospheric structure is fundamental also for the prediction
  of the observable properties, the emergent spectrum and the photometric colors.
  As shown by Bergeron, Saumon \& Wesemael (\cite{BSW}),
  when a WD with a pure hydrogen atmosphere cools down to T$_e\sim$ 5000 K and
  the molecular recombination begins, the collisionally induced absorption
  (CIA) of H$_2$-H$_2$ progressively becomes the most important opacity source 
  for the out-flowing infrared photons, so that the emergent spectrum departs from the blackbody
  appearance and the colors are blue shifted (see e.g. also Hansen \cite{hansen}).

  The present models take into account the liquid-solid phase transition 
  and the related energy release (see e.g. Paper I) but neglect the effect of 
  the phase separation during the crystallization of the core, 
  i.e. the chemical redistribution of carbon and oxygen due to
  the shape of the phase diagram for this binary mixture (Stevenson \cite{stevenson}).   
  We are interested in an evaluation of the relative effects on WD evolution of different
  physical aspects rather than to provide a reference standard model. Since
  the effect on WD evolution of chemical fragmentation has been already discussed
  several times in literature (see e.g. Segretain et al. \cite{segretain};
  Salaris et al. \cite{salaris97}; Montgomery et al. \cite{montgomery};
  Isern et al. \cite{isern97}, \cite{isern00}) we do not further investigate
  this issue.
\subsection{Initial-to-final mass relation and initial mass function}
  In order to compute isochrones, the age of the progenitor 
  at the entrance of the cooling sequence must be added to the cooling time.
  This quantity depends on the mass and the chemical composition of the progenitor. 
  The relationship between the main sequence mass and the final WD mass is
  still uncertain; it is sensitive to many poorly known 
  phenomena, like the extention of the H-burning convective core, which affects
  the stellar lifetime, and the efficiency of the mass loss mechanism, particularly
  during the TP-AGB phase when the star loses the greatest fraction of its mass, 
  which affects the final mass. 
 
  Semiempirical initial-to-final mass relations are only available for the 
  chemical composition of the solar neighborhood and for the Magellanic Clouds 
  (see e.g. Weidemann \& Koester \cite{koester}; Weidemann \cite{weidemann87},
  \cite{weidemann}; Herwig \cite{herwig}). 
  For the typical composition of the halo, the long distance and the large age 
  limits a direct determination of the initial-to-final mass relation. 
  Thus, in spite of the great uncertainty, we are forced to use theoretical
  relations.
  Figure \ref{Mwd_Mms_Z} shows the initial-to-final mass relationship
  from the homogeneous set of stellar models of low and intermediate mass stars
  computed by Straniero et al. (\cite{straniero97}) and Dominguez et al. (\cite{dominguez99}),
  with the same stellar evolutionary code used to provide the present
  set of WD models. The mass loss during the AGB phase has been obtained by 
  best fitting the period-mass loss diagrams obtained by Whitelock et al. (\cite{whitelock}).
  In stars with initial mass lower than 3 M$_{\odot}$, an additional pre-AGB mass loss
  has been considered based on the Reimers formula with the parameter
  $\eta=0.4$. Table \ref{MiMfTAB} reports the initial-to-final mass relations 
  shown in figure \ref{Mwd_Mms_Z}.   
\begin{table*}
\caption{Initial-to-final mass relationship for the four labeled
                 metallicities from our stellar models. See text for
                 details. Masses are in M$_{\odot}$}
\begin{center}
\begin{tabular}{c c c c c c c c c c c c c c c c c c}
\hline
\hline
\\
	M$_{MS}$ &       0.8 & 0.9 & 1.0 & 1.1 & 1.2 & 1.5 & 1.8 &
      2.0 & 2.3 & 2.5 & 2.7 & 3.0 & 4.0 & 5.0 & 6.0 & 7.0 & 8.0 \\
\\
\hline
\hline
\\
& & & & & & & &Z=0.02\\
\\
        M$_{WD}$  &      0.55 & 0.56 & 0.565 & 0.57 & 0.58 & 0.59 & 0.60 &
      0.61 & 0.62 & 0.63 & 0.64 & 0.67 & 0.81 & 0.91 & 0.985 & 1.06 & 1.14 \\
\\
& & & & & & & &Z=0.006\\
\\
       M$_{WD}$  &      0.55 & 0.56 & 0.565 & 0.57 & 0.58 & 0.59 & 0.60 &
      0.61 & 0.62 & 0.64 & 0.67 & 0.80 & 0.895 & 0.96 & 1.04 & 1.14 & 1.21 \\
\\
& & & & & & & &Z=0.001\\
\\
       M$_{WD}$  &      0.56 & 0.57 & 0.575 & 0.58 & 0.59 & 0.61 & 0.62 &
      0.64 & 0.68 & 0.72 & 0.76 & 0.82 & 0.90 & 0.98 & 1.05 & 1.15 & 1.22 \\
\\
& & & & & & & &Z=0.0001\\
\\
       M$_{WD}$  &      0.575 & 0.59 & 0.595 & 0.605 & 0.615 & 0.625 & 0.63 &
      0.66 & 0.72 & 0.77 & 0.815 & 0.86 & 0.92 & 0.99 & 1.09 & 1.18 & 1.26 \\
\hline
\hline
\label{MiMfTAB}
\end{tabular}
\end{center}
\end{table*}
  Three distinct regions with different slopes may be distinguished.
  Concerning the solar metallicity, these three regions are:
  M$_{i}<$ 3 M$_{\odot}$, 3 M$_{\odot} <$ M$_{i}$ $<$ 4.5 M$_{\odot}$ and 
  M$_{i}$ $> $ 4.5 M$_{\odot}$, respectively. At lower 
  metallicity the limit in mass of these three regions decreases, but the general features 
  are conserved. For all the three regions the adopted mass loss prescription 
  during the TP-AGB plays a crucial role in the determination 
  of the final WD mass. 
  Stars belonging to the first interval have a radiative core during the main sequence, 
  or develop a small central convective zone ($<<$ 0.5 M$_{\odot}$).
  Thus the mass of the hydrogen exhausted core (M$_H$) at the first thermal pulse
  depends on the efficiency of the H-burning shell operating 
  during the RGB phase and the following He-burning phase, and on the 
  corresponding evolutionary time scales.
  These parameters are very similar for all these low 
  mass stars so that their M$_H$ at the first thermal pulse
  is practically the same ($\sim 0.55$  M$_{\odot}$, see e.g. 
  Fig. 15 of Dominguez et al. \cite{dominguez99}).
  Then, the  most important phenomenon that influences the final WD mass 
  of these stars is the mass loss, taking place during the RGB and AGB phases,
  which erodes the envelope and limits the further growth of M$_H$
  during the thermally pulsing AGB.
  At variance, stars belonging to the second zone develop a sizeable convective core
  when they burn the central H. Thus, M$_H$
  at the first thermal pulse is mainly determined by the maximum extension attained by
  the convective core during the main sequence. In this case, the pre-AGB mass 
  loss have a negligible effect on the final mass, while that in the TP-AGB phase 
  is, once again, crucial. 
  Similar considerations are valid for the stars belonging to the third region,
  but, in this case, the growth of the core is also limited by
  the occurrence of the II dredge-up, during the early-AGB phase. 
  During the TP-AGB phase the recursive occurrence of a  
  deep III dredge-up brings inward fresh hydrogen and puts backward the
  outer edge of the  hydrogen exhausted core, thus preventing a substantial increase 
  of M$_H$. For this reason, 
  the final mass is only slightly larger than M$_H$ 
  at the beginning of the TP-AGB phase. 
  As shown in Fig. 20 of Dominguez et al. (\cite{dominguez99}) there is a general agreement 
  among the available predictions for M$_H$ at the first thermal pulse as a function 
  of the initial mass. The most important consequence of the assumption of a sizeable 
  convective core overshoot is the decrease of the maximum mass able to produce a C-O
  WD (M$_{up}$, Bertelli, Bressan \& Chiosi \cite{bressan}). 
  In order to compute WD LFs and synthetic 
  color-magnitude diagrams (CMDs) we also need the initial mass function
  (IMF) for the progenitor stars.
   In the following, if not explicitely stated, we have adopted
  a single power law IMF (dN/dM$_i \propto$ M$_i^{-\alpha}$), with a Salpeter exponent ($\alpha$=2.3). 
  The effect of varying both the slope of the IMF and the initial-to-final mass relationship 
  will be discussed in section 7. 
\section{Carbon and oxygen abundances in the core}
   \begin{figure}
   \centering
   \includegraphics[width=\columnwidth]{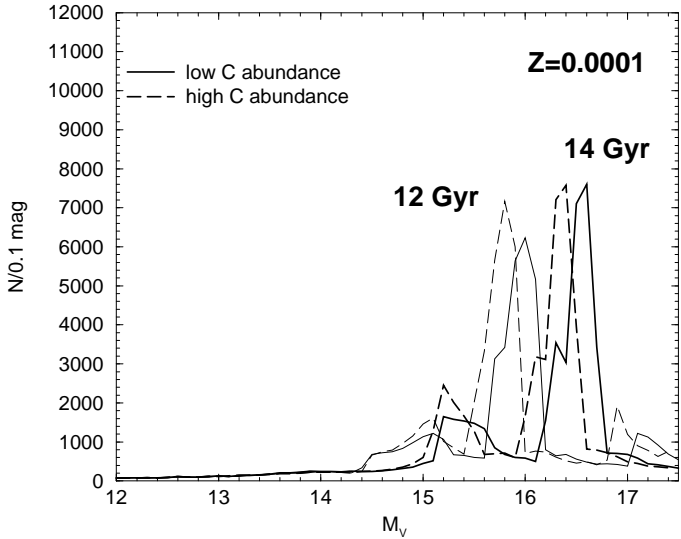}
   \caption{Theoretical LFs for 12 (thin lines)
  and 14 Gyr (thick lines) for WDs with low (solid lines) and
  high (dashed lines) C abundance.}
    \label{LF_CO}
    \end{figure}
  The thermal energy of a cool WD is almost entirely
  stored in the carbon and oxygen ions of the core.
  The cooling rate
  depends sensitively on the relative amount of carbon and oxygen.
  Carbon has an higher specific heat than oxygen, simply
  because the number of ions in a gram of C is larger.
  Thus, the cooling rate depends sensitively on the C/O ratio:
  the higher the carbon abundance in the
  core, the slower the cooling of the WD (Salaris et al. \cite{salaris97}; 
  Chabrier et al. \cite{chabrier}; Paper I). In addition, since  
  carbon crystallizes later than oxygen, the release of the latent heat
  and the onset of convective coupling are delayed when carbon is overabundant.
  
  The chemical abundance profiles in the core of a C-O WD are the result of
  the helium burning that occurred in the core and in the shell of the 
  progenitor star.
  As is well known, initially the 3$\alpha$ reaction produces carbon and 
  afterwards the $^{12}${C}$(\alpha, \gamma)^{16}${O} synthesizes oxygen. 
  The final chemical stratification in the core of a WD
  depends on the relative efficiency 
  of these two reactions. In particular, the reaction rate of the 
  $^{12}${C}$(\alpha, \gamma)^{16}${O} at an energy relevant for the stellar He-burning
  (about 300 KeV) is only known within a factor of 2 (see e.g. Buchman
  \cite{buchman}, Kunz et al. \cite{kunz}). The treatment of the convective 
  mixing during the core He-burning phase also affects the predicted C/O ratio
  (see Imbriani et al. \cite{imbriani}, Straniero et al. \cite{straniero}). 

  In Paper I we analyzed the changes of the cooling timescale
  of a 0.6 M$_{\odot}$ CO WD caused by the variation of
  the $^{12}${C}$(\alpha, \gamma)^{16}${O} reaction rate, within the present
  error bars, and by different prescriptions for the core convection.
 
  In this section we extend that analysis to masses from 0.5 to 0.9 M$_{\odot}$
  and calculate the corresponding isochrones and LFs.
  Then, two sets of WD cooling tracks were obtained 
  for two different C-O chemical
  stratifications in the core, the low carbon abundance model with a central carbon mass
  fraction of about 0.2 and the high ones with 0.50, a range that 
  roughly represents the present
  uncertainty in the theoretical prediction. The 
  former value appears in better agreement with the abundance derived from the
  power spectra of variable WDs (Metcalfe, Winget \& Charbonneau \cite{metcalfe}),
  although the feasibility of this kind of measurement has been questioned 
  (Fontaine \& Brassard \cite{fontainebrassard}).
  
  Figure \ref{LF_CO} shows the 12 (thin lines)
  and 14 Gyr (thick lines) LFs for WDs with low (solid lines) and
  high (dashed lines) C abundance. As expected, due to the lower heat
  capacity of the carbon-poor material, the cooling time is shorter in these
  models than the cooling time of the carbon-rich models.
  As a consequence, the WD LF peak occurs at lower luminosity. 
  The effect
  of the present uncertainty on the chemical profile in the core is
   sizeable.

 Taking into account that the position of
  the WD LF peak is a very sensitive function of the age ($\Delta$M$_v$/$\Delta$Age
  $\sim$0.3 mag/Gyr for GCs ages), the poor determination of the C abundance
  in the core translates directly into an uncertainty of about 0.6 Gyr in the inferred
  GC age.
\section{Conductive opacity}
   \begin{figure}
   \centering
   \includegraphics[width=\columnwidth]{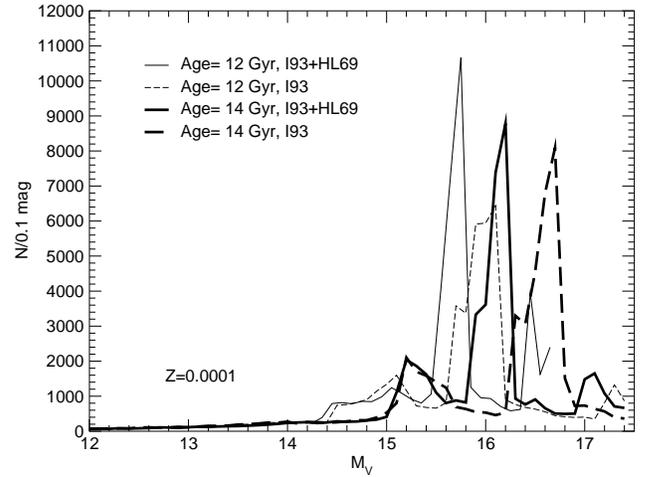}
   \caption{Theoretical LFs for 12 (thin lines) and 14 Gyr (thick
   lines) for WD models computed adopting the conductive opacity
   of Itoh and coworkers (I93) in the fully degenerate regime ($\theta=T_F/T <$ 0.1) and
   Hubbard \& Lampe (HL69) in the partially degenerate one ($\theta >$ 0.1)
   (I93+HL69, solid lines) and I93 in the whole structure (I93, dashed lines).}
    \label{LF_I93_HL69}
    \end{figure}
   \begin{figure}
   \centering
   \includegraphics[width=\columnwidth]{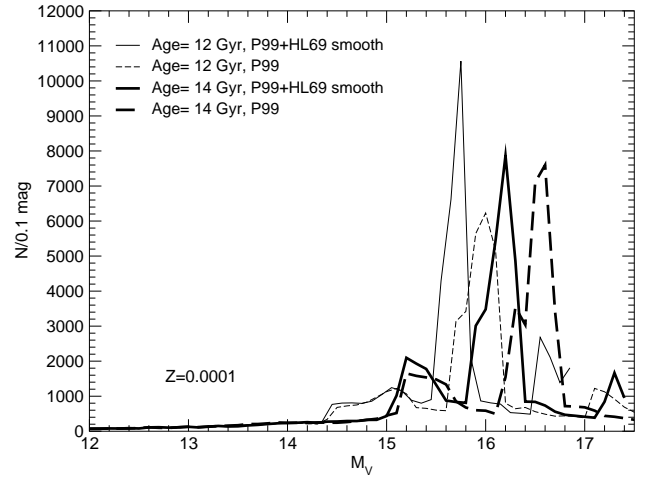}
   \caption{Theoretical LFs for 12 (thin lines) and 14 Gyr
   (thick lines) for WD models computed adopting the conductive opacity 
   of Potekhin (P99) and HL69 in the regions where, respectively, $\theta <$ 0.1
   and $\theta >$ 1, and a linear interpolation in the transition zone 
   (P99+HL69 smooth, solid lines) and the ones computed adopting the P99
   conductive opacity in the whole structure (dashed lines).}
   \label{LF_P99_HL69_P99}
   \end{figure}
   The electrons in the core of a WD are fully degenerate. In these conditions
   the energy transfer is largely dominated by electron conduction whose
   efficiency is high enough to keep the core almost isothermal.
   On the contrary, in the mantel (He-rich) and in the envelope (H-rich) the
   electrons are only partially or not degenerate, so that thermal conduction
   is less efficient. In bright WDs, the thin envelope is the most opaque region of
   the WD, a kind of insulating layer which regulates the temperature decrease
   of the core. The progressive development of the external convection 
   penetrating the partially degenerate region in cool WDs causes the already mentioned   
   convective coupling. Therefore, it is of primary importance for the description
   of the cooling process to understand the detailed treatment of the energy transport
   through the external layers (see e.g. D'Antona \& Mazzitelli \cite{dantona},
   Fontaine et al. \cite{fontaine}).

   In our knowledge, there are only two published conductive opacities suitable
   for partially degenerate regimes, the pioneering study by Hubbard \& Lampe
   (\cite{hubbard}, hereafter HL69) and Potekhin (\cite{potekhin},
   hereafter P99, see also Potekhin et al. \cite{potekhin2}) 
   \footnote{We have derived conductive opacity tables by means of the 
   Fortran codes provided by A.Y. Potekhin and available at the web site 
   (http://www.ioffe.rssi.ru/astro/conduct/conduct.html).}.
   Regarding the fully degenerate regime, in addition to the just mentioned works,
   there are also the fundamental contributions by Itoh and coworkers (Itoh et al.
   \cite{itoh83}, Mitake et al. \cite{mitake} and Itoh et al. \cite{itoh93},
   hereafter I93). 

   As discussed in detail in Paper I, the HL69,
   although nominally valid, should not be adopted in regions with a high Coulomb
   coupling parameter ($\Gamma >$ 10, where  $\Gamma$=(Ze)$^2$/k$_B$Ta and a
   is the interionic distance) due to an outdated treatment of the
   liquid-solid transition. On the other hand, the I93 computations are 
   strictly valid only in the completely
   degenerate regimes ($\theta$= T$_F$/T $<$ 0.1, where T$_F$ is the electronic Fermi 
   temperature), in fact they underestimate the
   contribution of the electron-electron interactions which is not negligible
   in the partially degenerate regimes. The only conductive opacity computations
   suitable for the whole WD structure and evolution are those by P99.

   In paper I we showed the dramatic effect on the evolution of a 0.6 M$_{\odot}$ WD
   of different assumptions for the conductive opacity in the thin envelope.
   At log $L/L_{\odot}$= -5.5 the adoption of the I93 in the whole WD structure
   leads to an age 17\% younger than that obtained by using HL69 in the 
   partially degenerate regime.
   In this section we extended
   that analysis to evaluate the effect on the WD isochrones and
   LFs of different assumptions on the conductive opacity.

   We have computed five sets of cooling tracks with masses in the range 0.5-0.9
   M$_{\odot}$ and the related set of isochrones and LFs.
   Figure \ref{LF_I93_HL69} shows the theoretical LFs for 12 (thin lines)
   and 14 Gyr (thick lines) for WD models computed adopting the conductive opacity
   of Itoh and coworkers (I93) in the fully degenerate regime ($\theta <$ 0.1) and
   Hubbard \& Lampe (HL69) in the partially degenerate one ($\theta >$ 0.1)
   (I93+HL69, solid lines) and I93 in the whole structure (I93, dashed lines).
   Notwithstanding that the two pairs of sets differ only in the conductive efficiency
   adopted in the partially degenerate regime, roughly corresponding to
   the thin external layer, whose mass is less than 1\% of the WD mass,
   the effect is very large causing a shift in the position of the LF peak
   of about 0.3 mag for 12 Gyr and 0.5 mag for 14 Gyr, which means roughly a difference
   in age of 1 Gyr and 1.6 Gyr, respectively. 

   Itoh suggested to use a different match between the I93 and the HL69
   conductive opacity (private communication). Thus, we have calculated an additional 
   set of cooling sequences, here referred to as I93+HL69 smooth, where 
   I93 is used for $\theta <$ 0.1, HL69 for $\theta >$ 1 and a linear interpolation
   is used between the two for 0.1 $<\theta <$ 1. 
   The variation of the WD LF peak is less than 0.1 mag.
   In addition, we have computed cooling models adopting the conductive
   opacity of Potekhin (P99) and HL69 in the regions where, respectively,
   $\theta <$ 0.1 and $\theta >$ 1, and a linear
   interpolation in the transition zone (here referred as P99+HL69 smooth). 
   To compute the conductive opacity of Potekhin 
   we use Fortran codes provided by A.Y. Potekhin. 
   The differences are small, as expected from a previous analysis presented in Paper I,
   because the P99 predictions are in good agreement with the I93 ones where this
   latter is nominally valid. 

   Figure \ref{LF_P99_HL69_P99} shows the LFs for 12 (thin lines) and 14 Gyr
  (thick lines) for the just presented "P99+HL69 smooth" WD models (solid lines)
  and the ones computed adopting the P99 conductive opacity in the whole structure
  (dashed lines). The discrepancy between the two sets of models is quite large,
  for the presented ages the shift of the WD LF peak is about 0.4 mag, roughly
  corresponding to 1.3 Gyr. As in the case shown in Fig. \ref{LF_I93_HL69},
  this result proves again the extreme sensitivity of the computed cooling
  times and thus of the theoretical WD LFs to the treatment of
  conductive opacity in the very thin (M$_{env}$/M$_{WD}<$0.01) envelope, where electrons
  are only partially degenerate. 
  To our knowledge, Potekhin 
  provides the only homogeneous conductive opacity computations suitable
  for the description of the whole WD structure available in the literature. 
  On the other hand, these results for nondegenerate plasmas, as stated by
  Potekhin himself, "are based on a 
  continuation from the degenerate domain (using Fermi-Dirac averaging) and can 
  be considered as order-of-magnitude estimates." (see e.g. description at 
  http://www.ioffe.rssi.ru/astro/conduct/conduct.html). 
  To provide firmer predictions of WD cooling times we need
  more precise treatments of electron conduction for partially degenerate regimes.

  At present the choice of the prescription adopted for the conductive
  opacity affects the age estimate of GCs at the level of 10\%.  
\section{Metal abundance}
   \begin{figure}
   \centering
   \includegraphics[width=\columnwidth]{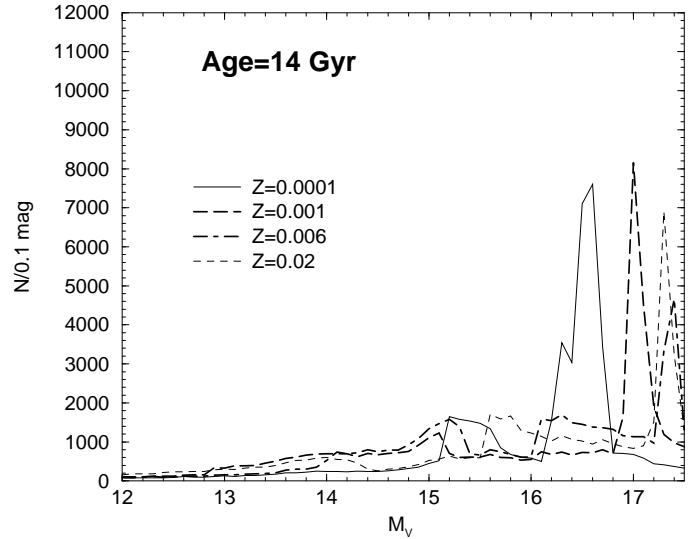}
   \caption{Theoretical LFs of 14 Gyr for the four labeled
 metallicities.}
    \label{LF_P99_Z}
   \end{figure}
    In Paper I we computed the evolution of two 0.6 M$_{\odot}$ WD models
    produced by progenitor stars with very different metallicity,
    respectively Z= 0.0001 and 0.02. The difference between the two
    cooling times are almost negligible (see e.g. Fig. 5 in Paper I),
    about 2\% at the faint end of our computations (log L/L$_{\odot}=$-5.5).

    The larger the metallicity, the thinner the external H-layer,
    because the H burning efficiency increases with a larger amount of CNO. 
    Due to the compact nature of these stellar remnants the
    gravitational settling is so efficient as to produce an envelope constituted by
    two layers of almost pure elements, i.e. a buffer of He surrounded by a very
    thin H surface envelope. 
    
    On the contrary, the WD isochrones and LFs are considerably
    affected by metallicity variations because, as is well known, the pre-WD evolutionary 
    time scale is a sensitive function of the original chemical abundance. 
    In addition, the initial-to-final mass relation
    changes with metallicity.
   
    In order to estimate the effect of the assumed chemical abundance,
    we computed four sets of isochrones and the related LFs for metallicities
    covering a range suitable for population I and II stars, i.e. Z=0.0001, 0.001,
    0.006 and 0.02. As already mentioned, we adopted the pre-WD evolutionary times and the
    initial-to-final mass relationship (see e.g. Fig. \ref{Mwd_Mms_Z}) from the homogeneous
    set of stellar models of low and intermediate mass stars
    by Dominguez et al. (\cite{dominguez99}) and Straniero et al. (\cite{straniero97}).

    Figure \ref{LF_P99_Z} shows the theoretical
    WD LFs for an age of 14 Gyr and the four labeled metallicities. 
    The position of the peak of the WD LF is a sensitive function 
    of the metal content of the progenitor stars. Between Z=0.0001 and Z=0.001,
    there is a difference in the visual magnitude of the peak of about 0.4 mag, 
    the same between Z=0.001 and Z=0.006. Thus, in order to use WDs to date GCs, 
    one should compute theoretical isochrones and LFs with 
    the cluster metallicity. For galactic GCs the typical error on [Fe/H] is of the 
    order of 0.1 dex. 
    This means that the typical uncertainty on the metallicity measurements only 
    slightly affects the estimates of the GC ages based on the WD cooling sequences,
    provided that WD isochrones 
    and LFs have been computed with the nominal value of the cluster metal abundance.
   \begin{figure}
   \centering
   \includegraphics[width=\columnwidth]{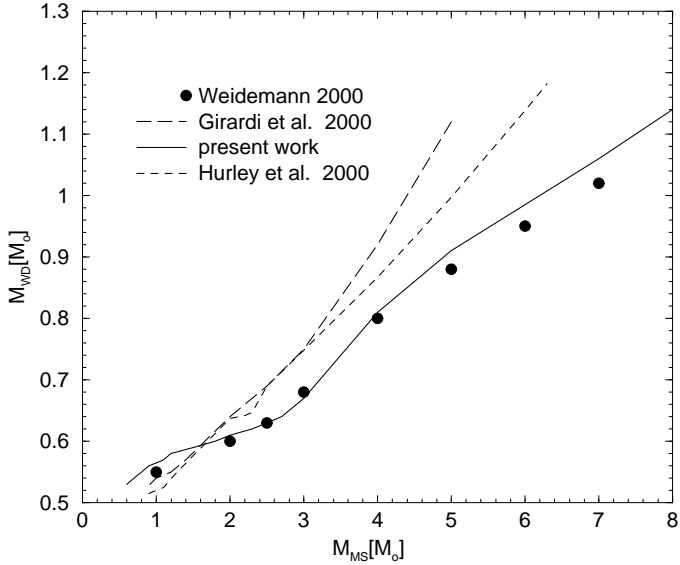}
   \caption{Initial-to-final mass relations for the solar chemical
    composition from: Weidemann (\cite{weidemann} filled circles); 
    Girardi et al. (\cite{girardi} dashed line); our relation
    (solid line); Hurley et al. (\cite{hurley00} short dashed line).}
    \label{Mi-Mms_autori}
   \end{figure}
\section{Initial-to-final mass relation and IMF}
   \begin{figure}
   \centering
   \includegraphics[width=\columnwidth]{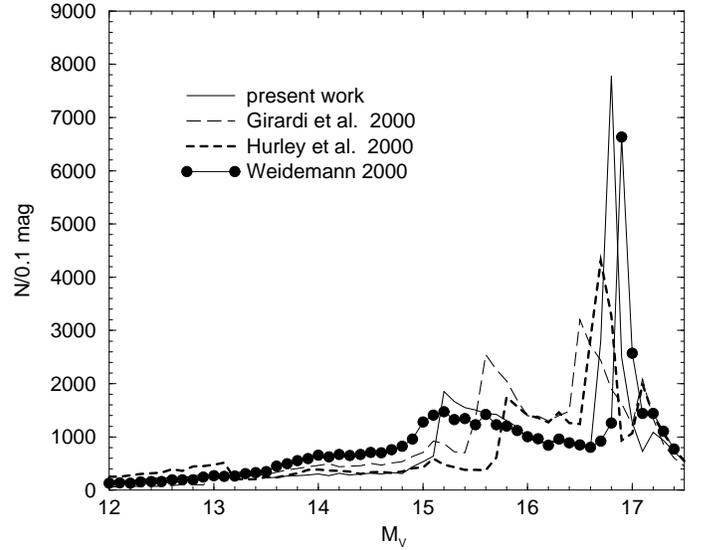}
   \caption{Theoretical WD LFs for 12 Gyr and different 
    initial-to-final mass relations for the solar chemical
    composition from: Weidemann (\cite{weidemann} filled circles); 
    Girardi et al. (\cite{girardi} dashed line); our relation
    (solid line); Hurley et al. (\cite{hurley00} short dashed line).}
    \label{LF_Mi-Mms}
   \end{figure}
   Figure \ref{Mi-Mms_autori} shows some of the most recent 
   initial-to-final mass relations for the chemical 
   composition typical of the solar neighborhood. 
   Weidemann (\cite{weidemann}, filled circles) derived the WD mass
   by comparing theoretical WD evolutionary models with the surface gravities and
   effective temperatures inferred from spectroscopy. Then the 
   initial mass is obtained by comparing the cluster turn off luminosity
   with stellar models of different masses 
    (see e.g. Weidemann \& Koester \cite{koester});
   thus also the semiempirical method depends 
   on some theoretical assumptions, like the convective core overshoot.    
   As mentioned above and as can be seen in Fig. \ref{Mi-Mms_autori},
   the initial-to-final mass relation is still quite uncertain, 
   even in the case of typical solar chemical composition.

   Girardi et al. (\cite{girardi}) rely on synthetic AGB models.
   This means that the evolution of the AGB stellar properties
   (the core mass, the total mass, the luminosity or the effective 
   temperature) is obtained by semiempirical prescriptions, 
   instead of solving the stellar structure equations
   (see e.g. Groenewegen \& de Jong \cite{groenewegen}; Bertelli et al. \cite{bertelli94};
   Marigo et al. \cite{marigo96}, \cite{marigo98}; Girardi \& Bertelli \cite{girardi98}).
   The Hurley et al. (\cite{hurley00}) relation has been computed with the 
   SSE package, kindly provided by Hurley, which uses 
   analytic formulae derived by fitting the stellar models
   computed by Pols et al. (\cite{pols98}).
   Our prescription for the initial-to-final mass relation is based 
   on full stellar evolution computations, from the pre-MS up to the AGB tip 
   (Straniero et al. \cite{straniero97}, Dominguez et al. \cite{dominguez99}, 
   Straniero et al. \cite{straniero05}). The pre-AGB mass loss is based on the 
   Reimers' formula ($\eta$=0.4), and an empirical calibration of the mass 
   loss-period relation is used for the AGB evolution (see Straniero et al. 
   \cite{straniero05} for details). Thus, the three curves reported in figure 
   \ref{Mi-Mms_autori} are representative of the different methods commonly
   used to derive the initial-to-final mass relation.
    
   Figure \ref{LF_Mi-Mms} shows the WD LFs related to the initial-to-final 
   mass relations described in Fig. \ref{Mi-Mms_autori}.
   In all cases we have adopted the same set of cooling evolutionary models and 
   the same slope of the IMF. We derived the pre-WD ages 
   from the homogeneous set of stellar models of low and intermediate mass stars
   by Straniero et al. (\cite{straniero97}) and Dominguez et
   al. (\cite{dominguez99}), except for the Girardi et al. (\cite{girardi})
   and the Hurley et al. (\cite{hurley00}) relations, for which we used the
   life times provided by their own stellar models.      
   The substitution of our relation by the Weidemann (\cite{weidemann}) one
   produces a shift of the peak of about 0.1 mag, 
   roughly corresponding to 0.3 Gyr. The same shift, but in the opposite 
   direction, is produced by substituting our relation with the 
   Hurley et al. (\cite{hurley00}) one. A larger effect is determined by 
   changing our relation with the Girardi et al. (\cite{girardi}) one; 
   in this case the peak shifts by about 0.3 mag, which translates into an age 
   error of 1 Gyr.    
   \begin{figure}
   \centering
   \includegraphics[width=\columnwidth]{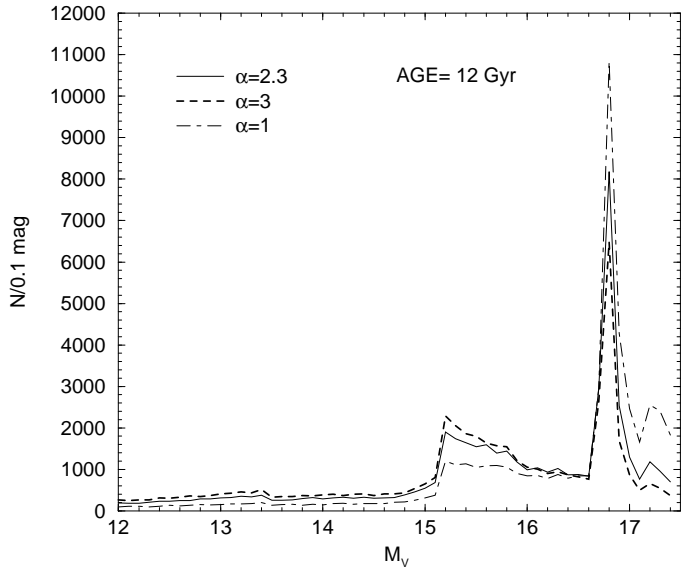}
   \caption{Theoretical WD LFs for 12 Gyr and solar metallicity. 
    Three different slopes of the IMF ($\propto$ M$^{-\alpha}$) 
    have been adopted: ${\alpha}$=2.3 (solid line); ${\alpha}$= 3
    (dashed line); ${\alpha}$= 1 (dot-dashed line).}
    \label{LF_IMF}
   \end{figure}  
   To illustrate the effect of a different mass distribution of the progenitor stars, 
   we have computed WD LFs adopting a single power law IMF 
   (dN/dM$_i \propto$M$_i^{-\alpha}$) varying the slope within a wide range, 
   i.e. ${\alpha}$= 1, 2.3 and 3. Figure \ref{LF_IMF} shows the 
   results; as expected the variation of the slope of the IMF affects only 
   the height of the peak, but leaves its position practically unchanged.
\section{Conclusions}
   WD cosmochronology is becoming a feasible 
   tool for dating stellar clusters. In this paper
   we have analyzed the accuracy of the
   available calibration of the WDs age, by 
   computing various sets of WD isochrones and LFs under 
   different prescriptions for the theoretical ingredients.    
   We discussed the effect on the WD LFs, and thus on 
   the globular cluster age derived with these compact objects, of 
   the assumed conductive opacity, C-O abundance in the core, 
   and initial-to-final mass relation and IMF. 

   The main effect is caused by the adopted conductive 
   opacity, in particular in the partially degenerate regime 
   characteristic of the thin outer layer. 
   For the range of ages typical of galactic GCs 
   the conductive opacity  affects the dating by 
   the order of 10\%. 

   Concerning the C-O amount in the core, the uncertainty in the age 
   is less than 5\%. Moreover, this contribution 
   will further decrease in the near future thanks to  
   experiments to 
   measure the  $^{12}{C}(\alpha, \gamma)^{16}{O}$ reaction rate
   at the energy of interest.

   The IMF does not affect the position of the peak and consequently 
   the estimated ages. 
   At variance, the initial-to-final mass relation plays an 
   important role in dating globular clusters because 
   it significantly affects the shape of the WD LF, in particular 
   the position of the peak. The uncertainty on the age can 
   reach 8\%. 
   
   Finally, given the sensitive dependence of the evolution of 
   the WD progenitors on the metal content, in order to use WDs to date GCs,
   one has to compute WD isochrones and LFs for the metallicity of the cluster. 
   In fact, between Z=0.0001 and 0.001 and between Z=0.001 and 0.006, 
   the difference in the inferred age is, for the range of interest, 
   of the order of 10\%. 

\begin{acknowledgements}
   We thank P. Bergeron for kindly providing us the model atmosphere 
   and N. Itoh and S. Shore for the many helpful comments. We are grateful to 
   J. R. Hurley for kindly providing us the SSE package.
\end{acknowledgements}

\end{document}